# Deflection of Interferometry Beams due to Transverse Refractive Index Gradient in SST-1

**Rajwinder Kaur and Asha Adhiya**[*]

Institute for Plasma Research, Bhat, Gandhinagar-382428, India

[*] ashaadhiya.ipr@gmail.com

**Abstract** Far Infrared interferometry is the main diagnostics method for electron density measurements in medium sized tokamaks. The transverse density gradients of plasma produce refractive effect, and the probing radiation does not propagate along a straight line and gets deflected. Therefore, it is necessary to evaluate the plasma effect on beam direction to verify if the used wavelength is compatible with the machine geometry. This paper presents the analytical results of refractive bending of THz beam due to transverse density gradients for circular and elliptical cross section plasma. The results from ray tracing integration to calculate refractive bending in D-shaped plasmas are also presented.

**Keywords:** Plasma diagnostics · D-shaped plasma · Transverse gradient · Refraction · Ray tracing

## 1 Introduction

The electron density profile, being directly related to both equilibrium and stability, is crucially required to understand the plasma behaviour in tokamaks. Plasma interferometry is the principle diagnostic method for electron density measurements, because it gives both spatial distribution and temporal variation with reasonable resolution. The principle of interferometric measurements is based on measuring the phase shift experienced by an electromagnetic wave crossing a plasma column [1,2]. Since the phase shift is a path integrated value, many channels are required to invert data and recover the density profile. This can be achieved by using a multichord interferometer to sample plasma simultaneously at several points.

One of the main problems in designing an interferometer is the selection of probing wavelength. The microwave interferometers which were commonly used in earlier medium sized tokamaks were limited to operating when $\nabla n_e L < 10^{14}$ cm$^{-3}$ [3] due to beam refraction effects, where L is the length of beam path and is the spatial gradient of electron density perpendicular to beam direction. This refraction may either result in the cross talk between adjacent channels or, even worse, may cause the signal to get shifted away from detector and get weakened to the point where it disappears completely. For circular and elliptical cross section plasma, analytical line integration of refractive bending of Gaussian beam along the beam path gives net refraction. The beam deflection has been calculated by analytical or numerical integration [4] of Snell's law or by detailed ray tracing calculations [5] in circular plasma.

A multichannel far infrared interferometer [6] has been developed to measure the spatial density distribution in Steady State Superconducting tokamak (SST-1) [7,8], at Institute for Plasma Research in India. Plasma in SST-1 tokamak, has been designed to have D-shaped cross-section resulting from the efforts to increase the energy confinement. The refractive effects produced become even more important for the plasma with D-shaped cross-section due to sharper density gradient. Therefore, in case of D-shaped plasma, more accurate integration of ray tracing equations, in the framework of Eikonal approximation, should be done to calculate the refraction.

This paper presents the results of line integration of refractive bending of interferometry beams through circular and elliptical cross section plasma. Further, the result of refractive bending for perpendicular propagation in D-shaped SST-1 plasma is also presented. Section 3 describes analytical calculations for circular and elliptical plasma cross-section. Section 4 presents the ray tracing equations, and the procedure followed. The results from ray tracing integration to calculate the refractive bending in D-shaped plasmas are presented in section 5. Finally, the conclusion is presented in section 6.

## 2   Interferometric Lines of Sight

The complete knowledge of two-dimensional spatial density distribution function for D-shaped SST-1 plasma requires probing along large number of lines of sight. The D-shaped cross section is achieved by a double null open divertor configuration, which imposes severe constraints on the number, location and orientation of possible lines of sight. For probing the plasma along vertical chords, one needs to send laser beam of suitable wavelength into the plasma from bottom port and collect/reflect them from top port. Fig 1(a) shows the vertical line of sight at radial location X = −0.025 m (R = 1.075). For lateral viewing, the probe beams of suitable wavelength enter the plasma column through a radial port and are reflected by the return retroreflectors mounted on the inner wall of the SST-1 vacuum vessel. Fig 1(b) shows the five lateral lines of sight. The spacing between them is adjusted so as to avoid cross talk between adjacent channels. The pivot point of these beams is kept so as to reduce the window size.

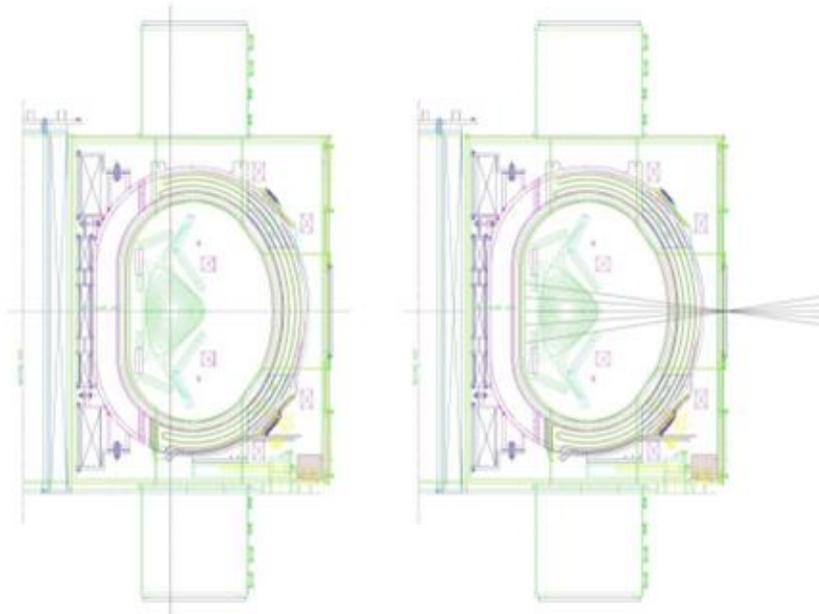

Fig 1:  Cross section of SST- 1 showing (a) Vertical path; (b) Lateral paths.

## 3 Beam Refraction Calculations using Analytical Methods

For ordinary mode propagation [9,10], the refractive index is given by

$$\mu_o = \sqrt{1 - \frac{\omega_p^2}{\omega^2}} \tag{1}$$

Plasma is transparent as long as the probing frequency $\omega$ is larger than the plasma frequency $\omega_p$.

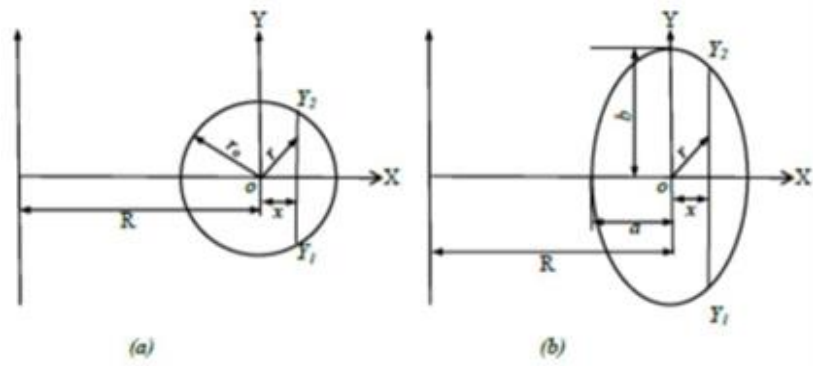

Fig. 2: Coordinate system for calculation of refraction in (a) circular plasma and (b) elliptical plasma

For $\omega < \omega_p$ the refractive index becomes imaginary, and the wave cannot propagate through the plasma. The density corresponding to $\omega = \sqrt{\frac{e^2 n_e}{\epsilon_o m_e}}$ is known as critical or cut off density. For $n_e \ll n_c$, the refractive index $\mu_o$, becomes

$$\mu_o = 1 - \frac{n_e}{2n_c} \tag{2}$$

For parabolic plasma density profile,

$$n(r) = n_o \left(1 - \frac{r^2}{r_o^2}\right) \tag{3}$$

The deflection angle is given by

$$\alpha(\mathbf{x}) = 2 \frac{n_e}{n_c} \frac{x}{r_o^2} \sqrt{r_o^2 - x^2} \tag{4}$$

For an axisymmetric parabolic profile, with the centre density $n_o$, the maximum of $\alpha(\mathbf{x})$ corresponding to $x = \frac{r_o}{\sqrt{2}}$ is $\alpha_m(\mathbf{x}) = \frac{n_e}{n_c}$.

For an elliptical cross section plasma, the deflection angle $\alpha(\mathbf{x})$ is

$$\alpha(\mathbf{x}) = 2\frac{n_e}{n_c}\frac{\kappa x}{r_o^2}\sqrt{r_o^2 - x^2} \tag{5}$$

where $\kappa = \frac{b}{a}$ is the ellipticity of the cross section, and $\alpha_m(\mathbf{x}) = \frac{\kappa n_e}{n_c}$ corresponding to $x = \frac{a}{\sqrt{2}}$

## 4 Ray Tracing Calculations

The ray tracing equations in the framework of Eikonal approximation in anisotropic dispersive plasma were derived by S. Weinberg [11]. The ray tracing involves the solution of these coupled differential equations with proper initial conditions on the ray initiation point and direction. The ray tracing starts with the probing beams entering the plasma with wave vectors $\mathbf{k}_R = 0$ and $\mathbf{k}_z = \mathbf{k}_o$ for vertical viewing. For lateral viewing, $\mathbf{k}_R$ and $\mathbf{k}_z$ are decided by the angle of the probing beam with the major radius and their entry point inside the plasma. Numerical integration of the ray equations is performed inside the plasma. This yields the ray trajectory along with the components of the wave vector $\mathbf{k}$ at each point on the trajectory. Along the trajectory, the wave must satisfy the real part of the dispersion relation

$$D(\mathbf{k}, \omega, \mathbf{r}, t) = 0$$

where we assume D is a slowly varying function of $\mathbf{r}$. Therefore, the trajectory satisfies two vector equations

$$\frac{d\mathbf{r}}{dt} = \frac{dD/d\mathbf{k}}{-dD/d\omega} \tag{6}$$

$$\frac{d\mathbf{k}}{dt} = \frac{-dD/d\mathbf{r}}{-dD/d\omega} \tag{7}$$

For ordinary mode propagation, dispersion relation is given by

$$D = k^2 c^2 + \omega_p^2 - \omega^2 \tag{8}$$

The four basic equations to be solved are

$$\frac{dR}{dt} = \frac{k_r c^2}{\omega} \tag{9}$$

$$\frac{dz}{dt} = \frac{k_z c^2}{\omega} \tag{10}$$

$$\frac{dk_R}{dt} = -\frac{e^2/\epsilon_o m_e}{2\omega}\frac{\partial n}{\partial R} \tag{11}$$

$$\frac{dk_z}{dt} = -\frac{e^2/\epsilon_o m_e}{2\omega}\frac{\partial n}{\partial z} \tag{12}$$

Therefore, we have four coupled ODEs to be solved for four unknowns $R, z, k_R, k_z$. Numerical integration of all four equations is done simultaneously inside the plasma. The flux surfaces for a typical D-shaped plasmas are used to model density as $n(R,z) = n_o(1 - \rho^2/a^2)^\gamma$, where $\gamma$ is the peaking parameter determining the sharpness of the density profiles. Values for $\gamma$ vary from 0.5 to 2.0 in SST-1 plasma. $\rho$ is described by the following two transcendental equations

$$R = R_o + \rho \cos(\theta + \delta \sin\theta)$$

$$z = \kappa\rho \sin\theta$$

where $R_o$ is the major radius such that $X = R - R_o$, $\delta$ is the triangularity and $\kappa$ is the ellipticity of the plasma cross section. For the ray tracing calculations $\kappa$ is taken as 1.7 and $\delta$ as 0.6.

The refractive bending of probing beam is evaluated by following the ray trajectory and calculating the deflection angle between the incident wave vector, $\boldsymbol{k_{in}}$, and the final wave vector, $\boldsymbol{k_f}$ as

$$\alpha = \cos^{-1}\left(\frac{k_{Rin}k_{Rf} + k_{zin}k_{zf}}{k_{in}k_f}\right) \qquad (13)$$

where $k_{Rin}$ and $k_{zin}$ are components of incident wave vector $\boldsymbol{k_{in}}$, $k_{Rf}$ and $k_{zf}$ are components of final wave vector $\boldsymbol{k_f}$. After exiting the plasma, the probing beam again travels along a straight line in the direction decided by the $\boldsymbol{k_f}$. The deviation of the beam at the exit window for vertical viewing is calculated accordingly. For lateral viewing the deviation is calculated at the return retroreflectors.

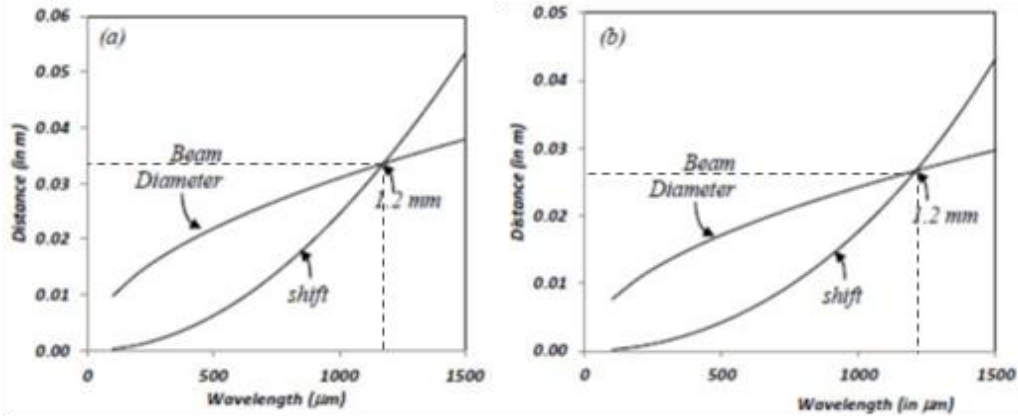

Fig 3: Comparison of beam diameter with shift due to maximum refraction as a function of wavelength (a) at window location for vertical viewing; (b) at reflectors on inside wall for lateral viewing

# 5 Results and Discussion

The refractive bending of probing beams results in translation of beams at the exit window for vertical viewing and at the return retroreflectors for the lateral viewing. The amount of shift depends upon various factors: the wavelength of probing radiation; the distance between two ports; the distance that wave travels inside plasma and the shape of the density distribution. One can think of changing the position of the incoming beam to compensate for the refraction. But, during the discharge, when electron density varies, the angle of refraction also varies accordingly, which makes this kind of compensation difficult. Hence the tendency is to limit the deflection up to a certain value so as to avoid the cross talk among adjacent channels. The reasonable upper limit to the refraction is set by not letting the displacement of beam exceed its diameter, d, at the first optical component after the beam exits from plasma.

Figure 3 shows the comparison between the maximum refraction and beam diameter as a function of probe wavelength for vertical and later*al* viewing. The upper limit of wavelength range for single pass vertical (Fig (a)) and double pass lateral (Fig (b)) viewing is $\lambda < 1.2$ mm.

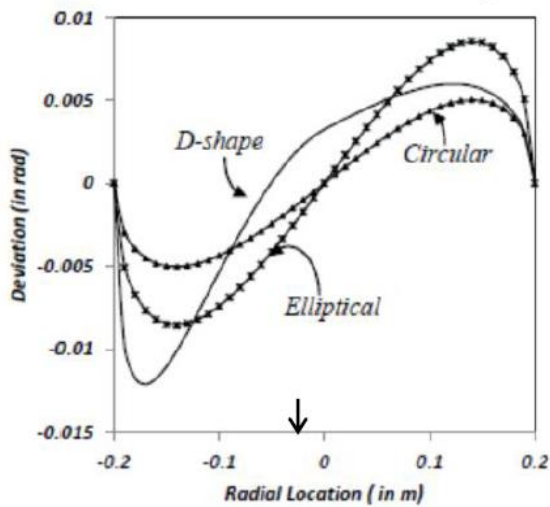

Fig 5: Variation of deviation of a 432.6 µm beam as a function of radial location for vertical paths for circular, elliptical and D-shaped plasma. Solid lines are the results from the ray tracing calculations and * are the points obtained from analytical calculations.

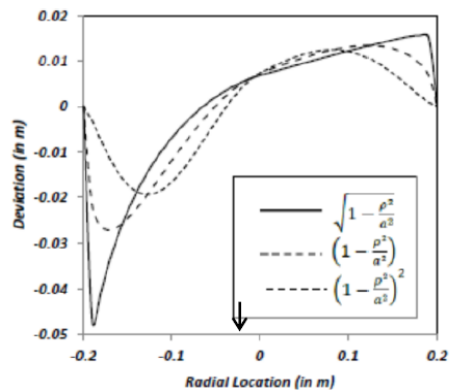

Fig 4: Variation of deviation of a 432.6 µm beam as a function of radial location for vertical paths

The lower limit of wavelength is set by the vibrations of optical components [10] which lead to spurious phase shifts in the interferometer measurements. The lower limit for vertical and lateral viewing for interferometer in SST-1 is $\lambda > 120$ μm [6]. The suitable source for vertical and lateral viewing interferometer for SST-1 is a 432.6 μm HCOOH laser or 337 μm HCN laser. Figure 4 shows the deviation of 432.6 μm vertical probing beam at the exit window as a function of the transverse position vector for SST-1 plasma with a central electron density of $3 \times 10^{19}$ m$^{-3}$ for three density profile shapes. The refraction of the beam increases with the peaking of density profile. Due to asymmetry of density gradients refraction is also asymmetric about minor axis in case of D-shaped plasma. The gradients are sharper on the inboard side of the plasma; therefore, refraction of the beam is also more. The location of the vertical probing beam is shown by an arrow at the bottom of the plot.

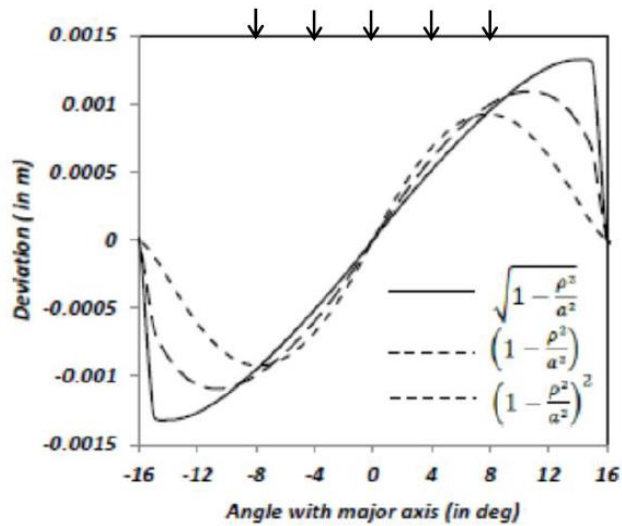

Fig 6: Variation of deviation of a 432.6 μm beam as a function of angle with major axis at the pivot point y tracing calculations and * are the points obtained from analytical calculations.

The validation of ray tracing calculations is done by reducing D-shape to circular ($\kappa = 1$, $\delta = 0$) and elliptical ($\kappa = 1.7$, $\delta = 0$) plasma and comparing the results with analytical calculations (Fig 5) for vertical viewing.

Figure 6 shows the deviation of 432.6 μm lateral probing beams at the return retroreflectors mounted on the inner wall of the vacuum vessel as a function of angle with major axis at the pivot point (shown in Fig 1). The up-down symmetry of the D-shaped plasma about the major axis results in the symmetrical refraction about the major axis. The lateral probe beams are shown in the figure by the arrows at the top.

# 6  Conclusion

The spatial resolution of an interferometer depends upon the probing wavelength. Therefore, it is necessary to select it very carefully. The refractive effect of plasma results in the deflection of probing beams at the optical components following their exit from the plasma. This results in the cross talk between adjacent channels, or even worse, shifts the signal out of detector's field of view. The upper limit to this deflection is set to be the beam diameter itself which, in turn, sets the upper limit to the probing wavelength. These calculations have been presented for D-shaped SST-1 plasma with the conclusion that the probing wavelength for vertical and lateral viewing of interferometer should not exceed 1.2 mm. The paper also summarises the results of refractive bending of 432.6 μm vertical and lateral probing beams in SST-1 plasma for various density profiles. These results are considered while designing the far infrared interferometer for SST-1.